\documentstyle[12pt,sw20lart]{article}

\input{tcilatex}
\begin{document}

\title{{\bf UNIFICATION OF BESSEL FUNCTIONS OF DIFFERENT ORDERS }}
\author{{\bf Mustapha.MEKHFI} \\
{\normalsize Laboratoire De Physique Mathematique,Es-senia 31100 Oran {\bf %
ALGERIE} }\\
{\normalsize And}\\
{\normalsize The {\bf Abdus salam }}\\
{\normalsize International Centre For Theoretical Physics ,34100 Trieste,%
{\bf ITALY}}}
\date{19 june 1999}
\maketitle

\begin{abstract}
We investigate the internal space of Bessel functions which is associated to
the group Z of positive and negative integers defining their orders.As a
result we propose and prove a new unifying formula ( to be added to the huge
literature on Bessel functions ) generating Bessel functions of real orders
out of integer order one's.The unifying formula is expected to be of great
use in applied mathematics.Some applications of the formula are given for
illustration

\bigskip

{\em Published in International Journal of Theoretical Physics }

{\em IJTP vol,39,No,4,2000}

{\footnotesize mekhfi@hotmail.com}
\end{abstract}

\newpage Very early studies [e.g., \cite{Truesdell},\cite{Infeld}] proposed
a unifying scheme for special functions showning that some of these
functions may originate from the same structure .For Bessel functons of
concern here their generating functions are representation ''states'' of the
derivative and /or integral operators of arbitrary orders.More precisely we
have the ''inner'' structure 
\begin{eqnarray}
m &\in &Z  \nonumber \\
\partial _{|m|} &=&\frac{\partial }{z\partial z}.\,\frac{\partial }{%
z\partial z}..........\frac{\partial }{z\partial z}.  \label{eq:1} \\
\partial _{-|m|} &=&\int zdz.\,\int zdz..............\int zdz.  \nonumber \\
\partial _{m}\Phi (z,t) &=&(-t)^{-m}\,\,\Phi (z)\,\,,m\epsilon Z  \nonumber
\\
\Phi (z,t) &=&\sum_{n=-\infty }^{n=\infty }\phi _{n}(z)\ t^{n}  \nonumber
\end{eqnarray}
\noindent Where $\int dz$ is the '' truncated '' primitive ,ie in defining
the integral we omit the constant of integration $\int \frac{df}{dz}dz=f$ .
For the polynomials such as Hermite and Laguerre for instance ,the
generating functions only involve the realization of the set N of positive
integers,with slight modifications of the derivative operators to account
for the conventions used in defining these polynomials.It is important to
note that although this common structure only set up the z-dependance of the
generating functions ,it is the ''dynamical'' part of the scheme so to
say.The t dependence is simply set by imposing some given desired
properties.For Bessel functions we require a ''symmetry'' between positive
and negative indices that is $J_{-n}=(-1)^{n}J_{n}$ ,while for the
polynomials it is the natural property of orthonormality that is invoked.In
this paper we unify Bessel functions of integer orders with those of real
orders as a complement to the unification of special functions initiated in
the above cited papers and subsequent papers.To our knowledge there is no
such relation , between integer and real orders ,known to exist.Both
functions are usually independently defined either by their own differential
equations or by their expansions in the form of a series or integral .The
mechanism underlying the unification we propose is of a different nature
than the one described above,but makes however use of the operators $%
\partial _{m}\,\,m\epsilon Z$ which served to define Bessel functions of
integer orders.

Let us first introduce the method in a simple example to see how the
mechanism works to convert an integer into a real.Suppose we are given an
abstract state $\mid n\rangle \,\,n\epsilon Z$ and a set of raising $\Pi
(m),m>0$ and lowering $m<0$ operators .Then it is easy to show , given that
data that the state $\mid n+\lambda >$ is related to the state $\mid n>$
through the following formula ( Provided the involved series converges which
is the case here ).

\begin{equation}
\mid n+\lambda >=\exp (-\lambda \sum_{m\epsilon Z/(0)}(-1)^{m}\frac{\Pi (m)}{%
m})\mid n>  \label{eq:b2}
\end{equation}

\noindent Fourier transforming the $\mid n>$ state as 
\begin{equation}
\mid n>=\int_{-\pi }^{\pi }\frac{d\theta }{2\pi }e^{in\theta }\mid \theta >
\label{eq:b3}
\end{equation}
\noindent

Where the $\Pi ( m ) $ operators acts on the $\mid \theta > $ state by
simple multiplication by the factor $e^{in \theta}$ .We get

\begin{eqnarray}
\mid n+\lambda > &=&\int_{-\pi }^{\pi }\exp (-\lambda \sum_{m\epsilon
Z/(0)}(-1)^{m}\frac{\Pi (m)}{m})\,\,e^{in\theta }\mid \theta >\frac{d\theta 
}{2\pi }  \nonumber  \label{eq:b4} \\
&=&\int_{-\pi }^{\pi }e^{i(n+\lambda )\theta }\mid \theta >\frac{d\theta }{%
2\pi }
\end{eqnarray}
\noindent In deriving the last line use have been made of the known formula $%
\sum_{m=1}^{\infty }(-1)^{m}\frac{sinm\theta }{m}=-\frac{\theta }{2}%
,\,\,\,-\pi <\theta <\pi \,\,.$

It is essential that the $\mid n>$ states have the appropriate weights for
the Fourier components otherwise the series defining the transformation
simply diverges. Reduced Bessel functions $\phi _{n}(z)=\frac{J_{n}(z)}{z^{n}%
}$ fit into the scheme .We indeed have a set of raising and lowering
operators $\Pi (m)=(-1)^{m}\partial _{m}\,\,m\epsilon Z$ \cite{watson} \cite
{smirnov}

\begin{equation}
(-1)^m\frac{d^m}{(zdz)^m} \phi_n(z)=\phi_{n+m} (z) \hspace {5mm} n \epsilon Z
\label{eq:b1411}
\end{equation}
\noindent

and the involved series does converge as it has almost the same structure as
for the example above.We then propose and prove the unifying formula.

\begin{equation}
\frac{J_{n+\lambda }(z)}{z^{n+\lambda }}=\exp (-\lambda \sum_{m\epsilon
Z/(0)}(-1)^{m}\frac{\Pi (m)}{m})\,\,\frac{J_{n}(z)}{z^{n}}  \label{eq:b5}
\end{equation}
\noindent The opertors act on the $\phi _{n}$ as 
\begin{equation}
\Pi (m)\phi _{n}=\phi _{n+m}  \label{eq:b6}
\end{equation}
\noindent

To check the above formula ,we could naively express the Bessel functions $%
\phi _{n+m} $ as entire series and then perform the sum over m .This is
however a cumbersome procedure.A simple and illuminating way to proceed is
to use the integral representation of $\phi _n $ , the "analog "of the $\mid
n > $ state Fourier decomposition.

\begin{equation}
\phi _{n}(z)=\frac{1}{2\pi i}\int_{l_{0}}{(\frac{1}{2})}^{n}\tau
^{-n-1}exp(\tau -\frac{z^{2}}{4\tau })d\tau  \label{eq:b7}
\end{equation}
\noindent Where $\tau $ is a complex variable and $l_{0}$ is a positively
oriented closed path encircling the origin one time .Expanding the
exponential in ~\ref{eq:b5} as . 
\begin{eqnarray}
\phi _{n+\lambda }(z) &=&\exp (-\lambda \sum_{m\epsilon Z/(0)}(-1)^{m}\frac{%
\Pi (m)}{m})\phi _{n}(z)  \nonumber \\
&=&\sum_{p=0}^{\infty }\frac{(-\lambda )^{p}}{p!}(\sum_{m\epsilon
Z/(0)}(-1)^{m}\frac{\Pi (m)}{m})^{p}\,\,\phi _{n}(z).  \label{eq:b8}
\end{eqnarray}
\noindent And using the propery that $\Pi (m_{1}).........\Pi (m_{p})=\Pi
(m_{1}+.........m_{p})$,the term of order $\lambda ^{p}$ takes the form.

\begin{equation}
\frac{ ( -\lambda)^p}{p!} \sum_{m_1}.........\sum_{m_p} \frac{
(-1)^{m_1+......m_p}}{ m_1.......m_p} \phi_{ m_1+.......m_p}  \label{eq:b9}
\end{equation}
\noindent Inserting the $\phi_n$ integral representation into ~\ref{eq:b9}
we get . 
\begin{eqnarray}
\frac{1}{ 2\pi i} \frac{ (-\lambda)^p}{p!} \int_{l_0}
\sum_{m_1}....\sum_{m_p} \frac{ (- 1)^{m_1+...m_p}}{ m_1.....m_p} {\ (\frac{1%
}{2} )}^{m_1+...m_p} \tau ^{-( m_1+......m_p +n )-1} exp( \tau- \frac{ z^2}{%
4\tau} ) d\tau .  \nonumber \\
=\frac{1}{ 2\pi i} \frac{ (-\lambda^p}{p!} \int_{l_0} (\frac{1}{2} )^n \tau
^{ -n-1} ( \sum_m (-1)^m \frac{ (2\tau )^{-m}}{m})^p exp( \tau- \frac{ z^2}{%
4\tau} )  \label{eq:b10}
\end{eqnarray}
\noindent The relevant term to sum is . 
\begin{equation}
\sum_{m \epsilon Z/(0)} {( -1)}^m \frac{ (2\tau )^{-m}}{m}  \label{eq:b11}
\end{equation}
\noindent It is to be noted that the above series diverges on the whole
complex plane except on the circle centring the origin and with half unity
radius, on which it converges uniformally.It is only on that circle do we
have the right to commute the signs $\sum $ and $\int $ when we insert ~\ref
{eq:b7} into ~\ref{eq:b9} .Now the miracle happens .To make the series
converge we just deform the path $l_0$ down to the above circle as the
integrand in ~\ref{eq:b7} has only essential singularities at $\tau=0$ and $%
\tau=\infty$. To compute the relevant sum ,put $2\tau=e^{i\theta}$ ,the
series then converges to .

\begin{eqnarray}
\sum_{m \epsilon Z/(0)} ( -1)^m \frac{ (2\tau )^{-m}}{m} &=& -2i
\sum_{m=1}^{m=\infty} {(-1)}^m \frac{ sin m \theta}{ m }  \nonumber \\
&=& i \theta \,\,\, -\pi <\,\theta \, < \pi \\
&=& ln 2\tau  \nonumber  \label{eq:b12}
\end{eqnarray}
\noindent

Where the branch cut of the logarithm is taken along the negative real axis
as $\theta $ ranges from $-\pi$ to $\pi$ .Inserting the result into ~\ref
{eq:b10} ,the latter becomes.

\begin{equation}
\frac{1}{ 2\pi i} \frac{{(-\lambda )}^p}{p!} \int_l {\ (\frac{1}{2} )}^n
\tau ^{-( n +\lambda )-1} (ln 2\tau)^p exp( \tau- \frac{ z^2}{4\tau} ) d\tau
.  \label{eq:b13}
\end{equation}
\noindent Where the new positively oriented path l is now getting round the
cut lying on the negative real axis.The above expression in ~\ref{eq:b13} is
however nothing than the $\lambda ^p $ order term of the real order Bessel
function $\phi_{n+\lambda} $ when it is written in terms of its integral
representation. 
\begin{equation}
\phi_{n+\lambda}(z) = \frac{1}{2\pi i} \int _{l_0} {\ (\frac{1}{2})}%
^{n+\lambda} \tau ^{-(n+\lambda)-1} exp( \tau - \frac{z^2}{4\tau} ) d \tau
\label{eq:b14}
\end{equation}
\noindent

To see this ,rewrite $(2\tau )^{-\lambda }=\exp (-\lambda ln2\tau )$ and
take the $\lambda ^{p}$ order which is $\frac{1}{p!}(-\lambda )^{p}(ln2\tau
)^{p}$ .We thus have succeed to prove the proposed unifying formula.\newline
Now we propose two applications of this formula for illustration . The first
immediate application is this .Bessel functions enjoy the essential property

\begin{equation}
(-1)^m\frac{d^m}{(zdz)^m} \phi_n(z)=\phi_{n+m} (z) \hspace {5mm} n \epsilon Z
\label{eq:b141}
\end{equation}
\noindent This formula is easily generalized to $\phi_{n+\lambda }$ with $%
\lambda$ real since the operator $\frac{d^m}{(zdz)^m}$ appearing in ~\ref
{eq:b141}commutes with the operator defining $\phi_{n+\lambda }$ in ~\ref
{eq:b5}.The second application is the computation of the generating function
of the reduced real order Bessel function assuming that we know the
generating function of the reduced integer order Bessel function .That is we
want to indirectly compute

\begin{equation}
\sum_{n=-\infty }^{n=\infty }\phi _{n+\lambda }(z)\,t^{n}  \label{eq:b15}
\end{equation}
\noindent Where t is a complex parameter .To perform the sum we use the
unifying formula.\vspace{20mm} 
\begin{eqnarray}
\sum_{n=-\infty }^{n=\infty }\phi _{n+\lambda }(z)\,t^{n} &=&  \nonumber \\
&=&\sum_{n=-\infty }^{n=\infty }\exp (-\lambda \sum_{m\epsilon Z/(0)}(-1)^{m}%
\frac{\Pi (m)}{m})\phi _{n}t^{n}  \nonumber \\
&=&\exp (-\lambda \sum_{m\epsilon Z/(0)}(-1)^{m}\frac{\Pi (m)}{m})\Phi (z,t)
\label{eq:b16}
\end{eqnarray}
\noindent Where $\Phi (z,t)$ is the generating function for Bessel functions
of integer orders.That is $\Phi (z,t)=\sum_{n=-\infty }^{\infty }\phi
_{n}(z)t^{n}$ .The action of the $\Pi (m)$ operators are simply . 
\begin{equation}
\Pi (m)\Phi =t^{-m}\Phi  \label{eq:b17}
\end{equation}
\noindent Then the third line in ~\ref{eq:b16} involves the following sum we
got used to. 
\begin{equation}
\sum_{m\epsilon Z/(0)}(-1)^{m}\frac{t^{-m}}{m}  \label{eq:b18}
\end{equation}
\noindent This series is only convergent on the circle of radius unity ,so
that we put $t=e^{i\theta }$ and find. 
\begin{eqnarray}
\sum_{m\epsilon Z/(0)}(-1)^{m}\frac{e^{-im\theta }}{m} &=&i\theta \,\,-\pi
<\theta <\pi  \nonumber \\
&=&lnt  \label{eq:b19}
\end{eqnarray}
\noindent Putting this into ~\ref{eq:b16} we get the end result that the
generating function for $\frac{J_{n+\lambda }}{z^{n+\lambda }}$ as defined
above is shown to be $t^{-\lambda }$ times the generating function for
Bessel functions of integer orders.Our formula shows that this is true on
the cirle of radius unity except on the the point -1 . 
\begin{equation}
\sum_{n=-\infty }^{n=\infty }\phi _{n+\lambda }(z)\,t^{n}=t^{-\lambda
}\sum_{n=-\infty }^{n=\infty }\phi _{n}(z)\,t^{n}=t^{-\lambda }\exp (\frac{t%
}{2}-\frac{z^{2}}{2t}).  \label{eq:b20}
\end{equation}
\noindent Note that our formula only predict the result on the circle .If we
assume in addition that the series is regular on the complex plane except on
the branch cut on the negative axis,then the result we found is valid
outside the circle and cover the whole domain of regularity.The unifying
formula will however show its power in other interesting applications which
are under active investigations.Let us for comparaison recompute the series
using a direct method .We will use the integral representation of the
reduced Bessel function ,otherwise the computation is simply awful.

\begin{equation}
\sum_{n=-\infty }^{n=\infty }\phi _{n+\lambda }(z)\,t^{n}=\frac{t^{-\lambda }%
}{2\pi i}\sum_{n=-\infty }^{n=\infty }\int_{l}(\frac{t}{2\tau })^{(n+\lambda
)}\exp (\tau -\frac{z^{2}}{4\tau })\frac{d\tau }{\tau }  \label{eq:b21}
\end{equation}
\noindent Making the change of variable $\frac{t}{2\tau }=e^{i\theta }$.We
rewrite the above expression as

\begin{equation}
\sum_{n=-\infty }^{n=\infty }\phi _{n+\lambda }(z)\,t^{n}=\frac{t^{-\lambda }%
}{2\pi }\int_{-\pi }^{\pi }\sum_{n=-\infty }^{n=\infty }e^{in\theta
}e^{i\lambda \theta }\exp (\frac{t}{2}e^{-i\theta }-\frac{z^{2}}{2t}%
e^{i\theta }))d\theta  \label{eq:b22}
\end{equation}
\noindent And knowing that the involved series converges to the sum of Dirac
distributions. 
\begin{equation}
\sum_{n=-\infty }^{n=\infty }e^{in\theta }=\sum_{p=-\infty }^{p=\infty }2\pi
\delta (\theta -2\pi p)  \label{eq:b23}
\end{equation}
\noindent We get after insertion of the above result . 
\begin{eqnarray}
\sum_{n=-\infty }^{n=\infty }\phi _{n+\lambda }(z)\,t^{n} &=&t^{-\lambda
}\int_{-\pi }^{\pi }\sum_{p=-\infty }^{p=\infty }\delta (\theta -2\pi
p)e^{i\lambda \theta }\exp (\frac{t}{2}e^{-i\theta }-\frac{z^{2}}{2t}%
e^{i\theta })d\theta \\
&=&t^{-\lambda }\exp (\frac{t}{2}-\frac{z^{2}}{2t})  \nonumber
\label{eq:b24}
\end{eqnarray}
\noindent Note that only the p=0 term in the above sum contributes since the
path of integration encircles the origin only one time.Here again we get the
same result with the advantage however that the result is valid on the whole
complex t plane except the cut .This is because we can always deform the $%
\tau $ contour so to make $\mid \frac{t}{2t}\mid =1$ and hence to sum the
series.Let us end up with the following worth remarks.First ,we have shown
that Bessel functions of real orders can be obtained from integer orders
ones through the mechanism we described above,and from this point of view
integer order Bessel functions are more ''fundamental''.Second,there is some
very known functions which appear in mathematical physics but are simply
constructed from various combinations of the linairely independant $J_{p}(z)$
and $J_{-p}(z)$ , where p is real .These are Hankel $%
H_{p}^{1}(z),H_{p}^{2}(z)$ and Neumann functions $N_{p}(z)$ which are
expressed as 
\begin{eqnarray}
N_{p}(z) &=&\frac{J_{p}(z)cosp\pi -J_{-p}}{sinp\pi }  \nonumber \\
H^{1}(z) &=&J_{p}(z)+iN_{p}(z) \\
H^{2}(z) &=&J_{p}(z)-iN_{p}(z)  \nonumber  \label{eq:b25}
\end{eqnarray}
\noindent These functions are also concerned with the above unification but
to give the formula which relates integer orders to real one's is not a
straigthforward matter although they are simply linear combinations of
Bessel functions.The reason is that the unifying formula acts on $%
J_{n}(z)/z^{n}$ and not directly on $J_{n}(z)$ and therefore we cannot
naively apply it to the defining formulas ~\ref{eq:b24} with p integer .We
managed however to unify Neumann's functions of different orders using our
unifying formula but following an indirect way .The result will appear in a
forthcoming paper. \vspace{15mm}

{\bf Acknowledgement} I would like to thank the head of the high energy
section at the {\bf abdus salam }international centre for theorethical
physics Dr S.Randjbar -Daemi for the continuous help and encouragments
during my stay as guest scientist at the centre.

\end{document}